\documentclass[submitting]{nst}

\usepackage{ulem}
\usepackage{bm}
\usepackage{subfigure,dcolumn}
\usepackage[T2A,T1]{fontenc}
\usepackage[russian,english]{babel}
\usepackage{amsmath}
\usepackage{cancel}

\usepackage{listings}
\usepackage{soul}
\begin{document}

\title{Application of quantum machine learning using variational quantum classifier in accelerator physics}

\author{He-Xing Yin}\thanks{These authors contributed equally to this work.}
\affiliation{The Institute for Advanced Studies, Wuhan University, Wuhan 430072, China}
\author{Zhi-Yuan Hu}\thanks{These authors contributed equally to this work.}
\affiliation{The Institute for Advanced Studies, Wuhan University, Wuhan 430072, China}
\author{Huan-Huan Zeng}
\affiliation{School of Physics and Technology, Wuhan University, Wuhan 430072, China}
\author{Jia-Bao Guan}
\affiliation{The Institute for Advanced Studies, Wuhan University, Wuhan 430072, China}
\author{Ji-ke Wang}
\email[Corresponding author, ]{Ji-ke Wang, The Institute for Advanced Studies, Wuhan University, Wuhan 430072, China, Jike.Wang@whu.edu.cn}
\affiliation{The Institute for Advanced Studies, Wuhan University, Wuhan 430072, China}

\begin{abstract}
Quantum machine learning algorithms aim to take advantage of quantum computing to improve classical machine learning algorithms. In this paper, we have applied a quantum machine learning algorithm, the variational quantum classifier for the first time in accelerator physics. Specifically, we utilized the variational quantum classifier to evaluate the dynamic aperture of a diffraction-limited storage ring. It has been demonstrated that the variational quantum classifier can achieve good accuracy much faster than the classical artificial neural network, with the statistics of training samples increasing. And the accuracy of the variational quantum classifier is always higher than that of an artificial neural network, although they are very close when the statistics of training samples reach high. Furthermore, we have investigated the impact of noise on the variational quantum classifier, and found that the variational quantum classifier maintains robust performance even in the presence of noise.
\end{abstract}


\maketitle

\section{Introduction}

Machine learning (ML) techniques have played a significant role in accelerator physics and are widely used in the design and optimization process of storage ring lattices. It is extremely time-consuming for simulating nonlinear dynamics such as dynamic aperture and momentum aperture. Therefore, many accelerator physicists use ML techniques, such as Gaussian processes or neural networks, as substitutes for simulations to save significant computational time and resources\cite{bib:1, bib:2, bib:3}. Combining ML techniques with multi-objective genetic algorithms (MOGAs) can greatly reduce the consumption of computational time and resources and potentially lead to better solutions than traditional MOGAs, as demonstrated by recent research$\cite{bib:4, bib:5}$. Additionally, online optimization is critical for achieving the design performance of an accelerator, and a recent study has demonstrated that a ML-based stochastic algorithm can be significantly more efficient than other stochastic algorithms commonly used in the accelerator community$\cite{bib:6}$. ML techniques have also been utilized in various other accelerator-related research areas, such as anomaly detection, stabilization of source size, beam diagnostics, and RF cavity design\cite{bib:7, bib:8, bib:9, bib:10, bib:11}. Overall, the application of ML techniques in accelerator physics has shown great promise in improving the efficiency and result of optimization, leading to better designs and more efficient operation of accelerators.

Classical computers use bits as the basic units of computation, while quantum computing uses quantum bits (qubits) as the basic units. Quantum machine learning (QML) is an emerging research field that aims to use quantum computing to augment classical ML\cite{bib:13}. QML has been widely applied in various fields, including high energy physics, drug discovery, and chemical compound space\cite{bib:14, bib:15, bib:16}. Quantum support vector machines (QSVM)\cite{bib:17, bib:18} and the variational quantum classifier (VQC)\cite{bib:19} are widely used for classification tasks, they can approach or even exceed the performance of their classical ML counterparts under certain conditions\cite{bib:20, bib:21, bib:22, bib:23}. Convolutional neural networks (CNNs), which can capture image dependencies\cite{bib:24, bib:25} and automatically learn important features from them\cite{bib:26, bib:27}, are one of the most widely researched areas currently. With the same number of training parameters, the quantum convolutional neural network can converge faster and achieve higher accuracy compared to the classical CNNs\cite{bib:28}. Gao et al. proposed a QML algorithm based on generative models that has the capability to represent probability distributions better than classical generative models in some examples, and exhibits an exponential speedup in learning and inference\cite{bib:29}.

This paper showcases the application of QML in accelerator physics. Unlike the previous ML approaches that directly predict DA from magnetic lattice settings, our method follows this paper$\cite{bib:46}$, it predicts DA by learning charged particles’ motion stability. Specifically, we train a QML model to infer long-term stability from each particle’s short-term trajectory. The training data is built by tracking sparsely sampled initial positions in phase space. Once trained, the model classifies whether a particle will be lost or survive based solely on its early trajectory, enabling us to reconstruct the DA boundary. By classifying particle confinement status (lost or stable) through early trajectory patterns, the approach effectively converts DA boundary determination into a binary classification task. Considering that traditional DA evaluation requires tracking dense initial conditions over long periods, this method allows to reduce computational costs by tracking fewer initial conditions for similar durations. Preliminary, VQC’s results indicate that QML has potential for DA optimization. In future work, we will also explore other QML algorithms (e.g. VQE) to perform comprehensive lattice optimization, for example simultaneously taking DA and other objects (like emittance, life-time, etc.) as the optimization targets.

The paper is organized as follows. Section 2 provides a detailed description of the widely used QML algorithm, VQC. Section 3 outlines the optimization process setup. The results and discussion are presented in Section 4, followed by a summary of the conclusions in Section 5.

\section{Quantum Machine Learning Models}\label{sec:artwork}

\begin{figure}[!htb]
\includegraphics
  [width=0.9\hsize]
  {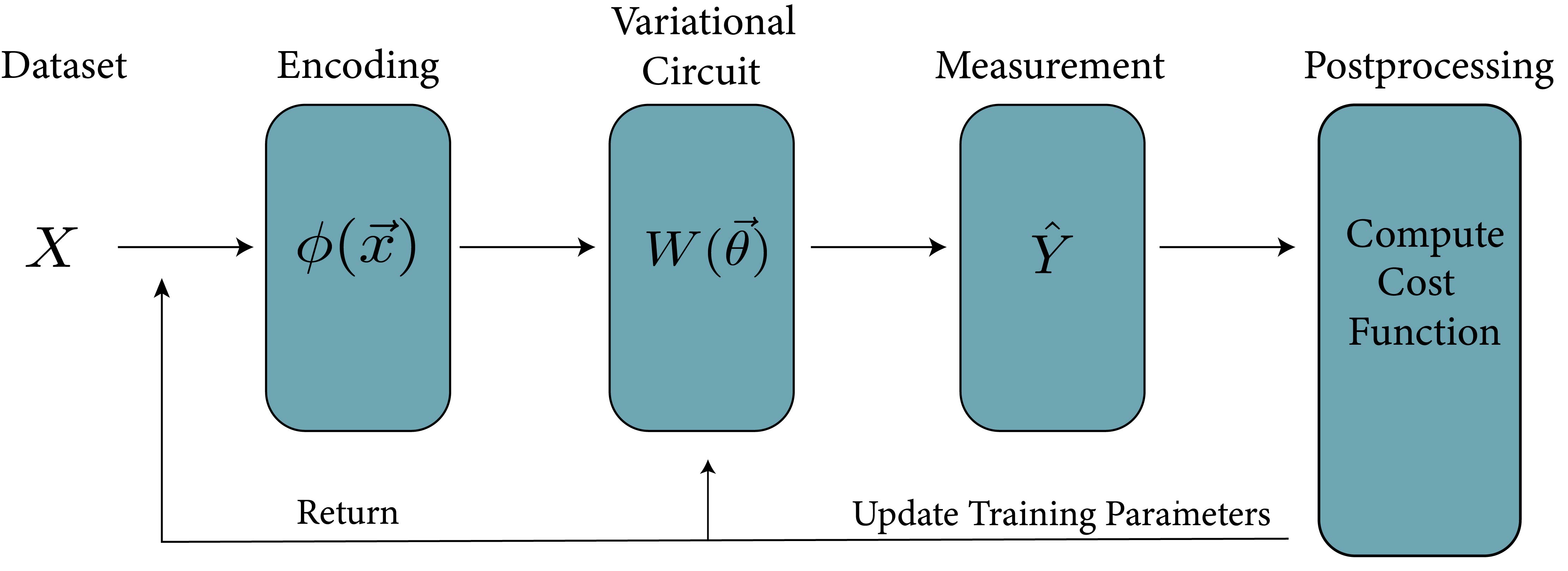}
\caption{Schematic diagram of the VQC model. $X$ represents the dataset, ${\Phi}\left(\vec{{x}}\right)$represents the quantum gate operation that maps classical data to the quantum state of qubits, ${W}\left(\vec{{\theta}}\right)$ represents the optimizable quantum computing layer consisting of quantum gates with adjustable parameters, which allows the parameters to be continuously adjusted during the training process to minimize the loss function. $\hat{{Y}}$ represents the classical data obtained after measuring the qubits in the final state of the quantum circuit.}
\label{fig:two-column-figure}
\end{figure}

A quantum circuit is a fundamental element of quantum algorithms and is formed by combinating qubits and quantum gates. Circuits that have variable parameters are known as parametrized quantum circuits (PQCs)\cite{bib:30}. VQC is a quantum neural network based on PQCs and has a structure similar to the classical artificial neural network (ANN). As illustrated in Fig.~\ref{fig:two-column-figure}, the VQC consists of three primary components: data encoding, variational circuit, and measurement and post-processing. 

Assuming that a classical dataset X has M samples, where each sample has N features represented by the feature vector $\vec{x}$. The output state of a quantum circuit for each feature vector can be expressed as

\begin{equation}\label{eq:1}
\left| \psi(\vec{x}; \vec{\theta}) \right\rangle = U_{{W}(\vec{\theta})} \, U_{{\phi}(\vec{x})} \, \left| 0 \right\rangle
\end{equation}

where $U_{{W}(\vec{\theta})}$ corresponds to the variational circuit and $U_{{\phi}(\vec{x})}$ corresponds to the data encoding, $\vec{\theta}$ are the training parameters of the variational circuit. The training process for the VQC model can be summarized as follows: first, classical data is encoded into the quantum system and then passed through the variational circuit. The output state of the variational circuit is measured, and then the training parameters of the variational circuit are updated based on the cost function derived from that state. This process is repeated until the training termination condition is met.

\subsection{Data Encoding}

Data representation is crucial in the application of ML algorithms, and the same is true for QML algorithms. The process of transforming classical data into quantum systems is known as data encoding, which is also called data embedding or loading\cite{bib:31, bib:32, bib:33}. In this paper, we present two widely used data encoding methods, amplitude encoding and angle encoding.

Amplitude encoding is a data encoding method used in QML algorithms, which maps classical data into the amplitude of a quantum state. Specifically, each feature of a classical data sample is represented by a quantum state amplitude. Thus, a data sample $\vec{x}$ in X becomes the amplitude of a quantum state $| \psi_{\vec{x}} \rangle$ by amplitude encoding,

\begin{equation}\label{eq:2}
| \psi_{\vec{x}} \rangle = \sum_{i=1}^{N} x_i | i \rangle
\end{equation}

where $x_i$ is the i-th feature of $\vec{x}$,  $| i \rangle$ is the i-th computational basis state. It should be noted that a quantum circuit with n qubits has a total of $2^n$ quantum states, which means that a data sample with N features only requires at least $\log_2 N$ qubits to be encoded into the quantum system. If the number of features is less than the number of amplitudes of the quantum state used for encoding, the extra amplitudes can be set to 0 to satisfy the normalization condition.

Angle encoding is another commonly used data encoding method in QML algorithms that maps classical data into rotation angles for the rotation gates in a quantum circuit. For example, the sample $\vec{x}$  by angle encoding can be encoded as

\begin{equation}\label{eq:4}
\left| \psi_{\vec{x}} \right\rangle = \bigotimes_{i=1}^{N} \left( \cos(x_i) \left| 0 \right\rangle + \sin(x_i) \left| 1 \right\rangle \right)
\end{equation}

In this paper, we use $R_x$ gates for angle encoding, which are single-qubit quantum gates that rotate the qubit at an angle of $\theta$ along the x-axis, and are denoted as

\begin{equation}\label{eq:5}
R_x(\theta) = \begin{pmatrix}
\cos\left(\frac{\theta}{2}\right) & -i\sin\left(\frac{\theta}{2}\right) \\
-i\sin\left(\frac{\theta}{2}\right) & \cos\left(\frac{\theta}{2}\right)
\end{pmatrix}
\end{equation}

\begin{figure}[!htb]
\includegraphics
  [width=1\hsize]
  {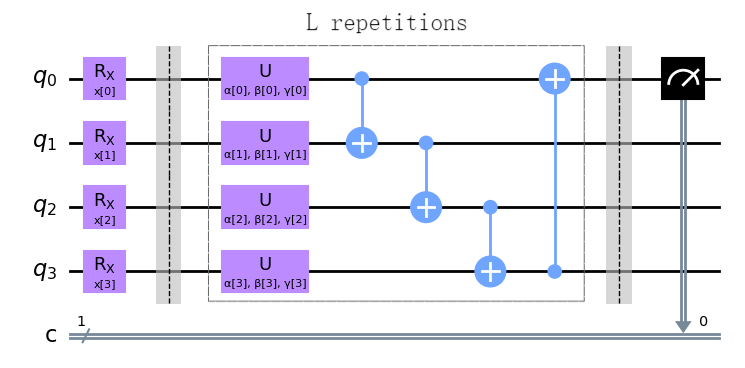}
\caption{Circuit structural diagram of a VQC model with angle encoding. L represents the number of repetitions of the parameterized layers, $R_x$ refers to the $R_x$ gate (rotation gate around the X-axis), and $\alpha$, $\beta$, and $\gamma$ are the angle parameters.}
\label{fig:three-column-figure}
\end{figure}

Angle encoding differs from amplitude encoding in that it requires at least N qubits to encode a classical data sample with N features. As shown in Fig.~\ref{fig:three-column-figure}, a circuit with 4 qubits can employ 4 $R_x$ gates to encode the four features of a sample into the quantum system using angle encoding.

\subsection{Variational circuit}
After encoding the classical data into the initial quantum state $\left| \psi_{\vec{x}} \right\rangle$, variational circuit $U_{{W}(\vec{\theta})}$ maps this state to another quantum state $\left| \psi(\vec{x}; \vec{\theta}) \right\rangle$. A variational circuit $U_{{W}(\vec{\theta})}$ can be composed of one or more PQCs, which can be expressed as 

\begin{equation}\label{eq:6}
U_{W(\vec{\theta})} = U_{W_k (\vec{\theta_k})} \cdots U_{W_i (\vec{\theta_i})} \cdots U_{W_2 (\vec{\theta_2})} U_{W_1 (\vec{\theta_1})}
\end{equation}

where $U_{W_i (\vec{\theta_i})}$ represents the i-th PQC. For a variational circuit, there are two important criteria: expressibility and entanglement ability\cite{bib:34}. The former characterizes the coverage of the circuit in the Hilbert space, while the latter represents the ability of the circuit to capture non-trivial correlations in the data. 

Fig.~\ref{fig:three-column-figure} depicts a strongly entangled layer of a variational circuit with strong expressibility and entanglement, represented in a square-shaped area. This type of circuit has been widely utilized and has shown good performance\cite{bib:20, bib:35}. The strongly entangled layer consists of generic single-qubit rotation gates with the same number of qubits, as well as a series of CNOT$\cite{bib:37}$ gates that entangle qubits in a circular topology. The generic single-qubit rotation gate with three angles is defined as follows

\begin{equation}\label{eq:7}
U(\alpha, \beta, \gamma) = \begin{pmatrix}
\cos\left(\frac{\alpha}{2}\right) & e^{-i\gamma} \sin\left(\frac{\alpha}{2}\right) \\
e^{i\beta} \sin\left(\frac{\alpha}{2}\right) & e^{i(\beta + \gamma)} \cos\left(\frac{\alpha}{2}\right)
\end{pmatrix}
\end{equation}

where $\alpha, \beta, \gamma$ represent the rotation angles along the three axes respectively. The variational circuits can enhance expressibility and entanglement by increasing the number of strongly entangled layers. However, there are several trade-offs that need to be considered\cite{bib:36}:

\begin{description}

\item[Speed] As the complexity of the circuit increases, the simulation speed of the quantum circuit decreases sharply. Therefore, increasing strongly entangled layers will slow down the algorithm.
\item[Performance] Insufficient strongly entangled layers may cause the algorithm to fail to find the optimal solution.
\item[Noise] Currently, most quantum computers are noisy intermediate-scale quantum computers (NISQs)$\cite{bib:38}$, which means that noise is unavoidable. The noise on NISQs becomes increasingly impactful on the performance of the algorithm as the number of the strongly entangled layers increases.
\end{description}

Furthermore, experimental results have shown that increasing the number of strongly entangled layers beyond a certain number on the ideal model may not lead to any further improvement in the performance of VQC\cite{bib:21}.

\subsection{Measurement and postprocessing}
Converting the information from the quantum state into classical data is a crucial step in the VQC model. As obtaining the output quantum state of a quantum circuit directly is an extremely challenging task, Qiskit\cite{bib:39} and IBM quantum computers\cite{bib:40} typically acquire the information of the quantum state indirectly by measuring each qubit of the circuit repeatedly. Each measurement of a qubit results in either $\left| 0 \right\rangle$ and $\left| 1 \right\rangle$. By measuring each qubit multiple times and analyzing the probability distribution of the occurrence of the two states $\left| 0 \right\rangle$ and $\left| 1 \right\rangle$, the output quantum state can be inferred.

For binary classification problems, many studies have only measured a single qubit\cite{bib:20, bib:21}, and we also only measured the first qubit as shown in Fig.~\ref{fig:three-column-figure}. Suppose a binary classification problem is labeled either 0 or 1,

\begin{equation}\label{eq:8}
\hat{y} = \begin{cases} 
0, & P(|0\rangle) \geq 0.5 \\
1, & P(|0\rangle) < 0.5
\end{cases}
\end{equation}

Where $\hat{y}$ is the prediction label and $P(|0\rangle)$ is the probability of occurrence of the state $\left| 0 \right\rangle$. The remaining post-processing steps and the specific details of the experimental setup will be described in detail in Section 3.

\section{Experiment setup}
\subsection{Data Description}

Due to the nonlinear effects of particle motion in a particle accelerator, there is a limitation beyond which the motion amplitude of particles will further diverge, leading to particle loss. This limitation is referred to as the dynamic aperture\cite{bib:41, bib:42}. For a particle to survive in the long-term, its initial position must not exceed the range of the dynamic aperture. The dynamic aperture can be described using a six-dimensional phase space $[x, x', y, y', z, \delta]$, where x,y represent the coordinates of particles in horizontal and vertical directions, $x',y'$ denote the divergence in horizontal and vertical directions, and $z,\delta$ represent the longitudinal position difference and momentum difference of the ideal particle. For a given lattice of a storage ring, the conventional method of computing the dynamic aperture involves long-term tracking. In this paper, long-term tracking refers to the process of tracking particles for 1000 turns. Long-term tracking is performed along a certain number of rays, starting from the inside and moving outward until the first particle loss position appears on each ray. The dynamic aperture is then determined by connecting the previous particle of the first lost particle on each ray. All particle tracking in this study was performed using the Accelerator Toolbox (AT)\cite{bib:43}.

The conventional method of computing the dynamic aperture through long-term tracking is computationally expensive. Therefore, ML techniques have gained widespread popularity in recent years for evaluating the dynamic aperture \cite{bib:1, bib:4, bib:44, bib:45, bib:46}. Most experiments that use ML techniques to evaluate the dynamic aperture utilize the magnetic lattice settings of the storage ring as features to predict the size of the dynamic aperture. However, since there is no clear dependency between these features and the dynamic aperture, the accuracy of ML predictions may decrease as the variable space changes. Although continuously retraining the model when new data is included can alleviate this limitation, it increases computational time and requires constant attention to the accuracy of dynamic aperture predictions when applied to new storage rings.

\begin{figure}[!htb]
\includegraphics
  [width=1\hsize]
  {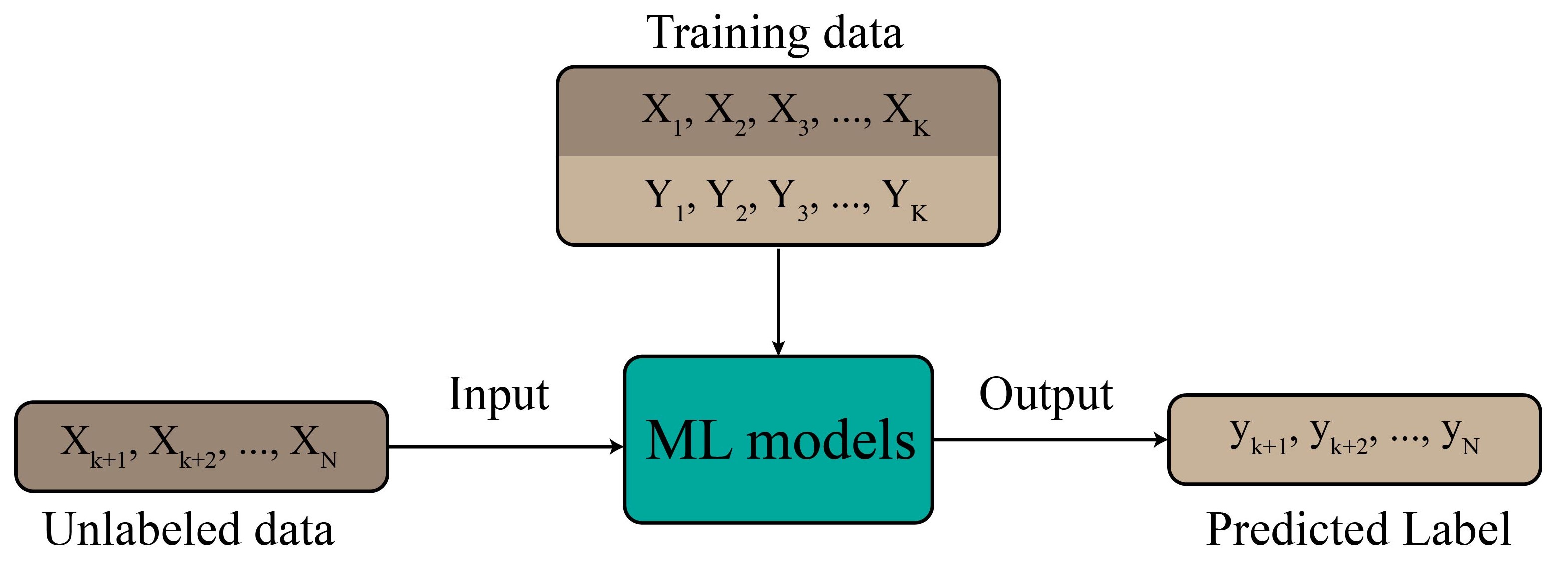}
\caption{Schematic diagram of using ML models to predict the dynamic aperture. N refers to the total number of particles used to distinguish the dynamic aperture. The trajectory of the i-th particle is represented by $X_i$, $Y_i$ denotes the label of the i-th particle obtained from long-term tracking, and $y_i$ represents the predicted label of the i-th particle acquired from the ML model.}
\label{fig:four-column-figure}
\end{figure}

Different from ML models that rely on magnetic lattice settings to predict dynamic aperture size, this paper follows the method introduced by Jiao et al. to directly predict the dynamic aperture profile by learning the motion stability of the particles\cite{bib:46}. Experiments conducted by Jiao et al. have demonstrated the high accuracy of this method, which remains effective even when the physical model of a storage ring is altered. As shown in Fig.~\ref{fig:four-column-figure}, we will describe how to use this method to obtain the dynamic aperture. A series of initial particles are uniformly generated within a large semicircle that can cover the dynamic aperture of a storage ring. Short-term tracking (few turns or even only one turn) is performed on all particles to obtain the trajectory of particles. We used 10\% or fewer particles for long-term tracking and treated these particles as the training set for ML, where the trajectories of short-term tracking are treated as features X (consists of $[x,x',y,y']$ of different positions of the storage ring obtained by short-term tracking), and the stability of long-term tracking is treated as label Y (0 or 1 for unstable or stable, respectively). The remaining unlabeled particles can be predicted to be stable or not by the trained ML model. After all the initial particles have been labeled, the dynamic aperture can be obtained based on the separation between the surviving particles and the lost particles.

\subsection{Optimization}
In training QML models as well as ANN models, the loss function is chosen to be cross entropy loss function\cite{bib:47}. The goal of the training process is to find the values of the training parameters for the model that minimizes the loss function L for a classification problem:

\begin{equation}\label{eq:9}
\mathcal{L} = \frac{1}{N} \sum_{i=1}^{N} \left( - Y_i \log p_i - (1 - Y_i) \log (1 - p_i) \right)
\end{equation}

where N represents the number of training samples, $Y_i$   represents the label of sample i, $p_i$ represents the probability that sample i is predicted to be $Y_i$ .

Gradient-based optimizers\cite{bib:48}, such as ADAM\cite{bib:49}, stochastic gradient descent (SGD)\cite{bib:50}, are commonly used to train ML models. Assuming a model has k training parameters, these optimizers need to evaluate the loss function 2k times in one iteration. However, simulating quantum circuits for QML models can be very time-consuming, making it challenging to evaluate the loss function multiple times during the training process. Since constrained optimization by linear approximations (COBYLA)\cite{bib:51} is a gradient-free optimizer that only needs to evaluate the loss function once per iteration, regardless of the number of training parameters, we use it as the training optimizer for the QML presented in this paper. In contrast, the ANN counterpart utilizes ADAM as the optimizer.

\begin{figure}[!htb]
\includegraphics
  [width=0.8\hsize]
  {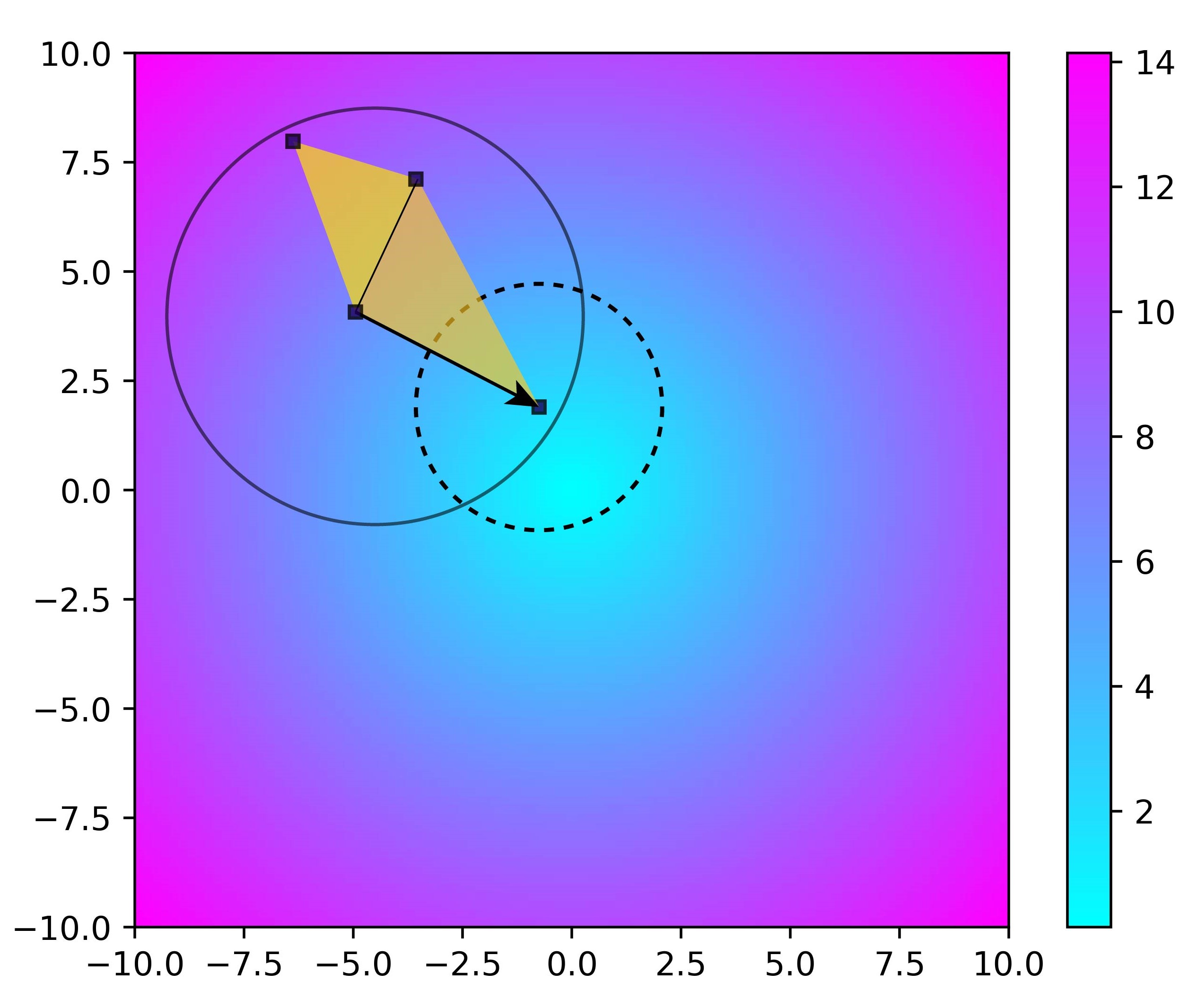}
\caption{The schematic diagram of COBYLA in a two-dimensional variable space. The smaller value corresponding to the colorbar represents the better the better objective value. The solid circle represents the current trust region, the triangle on the upper left indicates the initial simplex, and the arrow indicates the transition from existing vertices of the simplex to the best point in the current trust region. The dashed line indicates the new trust region formed based on the current best point.}
\label{fig:five-column-figure}
\end{figure}

As shown in Fig.~\ref{fig:five-column-figure}, COBYLA constructs a simplex consisting with m+1 points in the trust region, where m denotes the number of variables. In each iteration, a new point is generated using interpolation and linear approximations based on the vertices of the simplex. This point may replace an existing vertex on the simplex, either to improve the shape of the simplex or because it is the best point found so far. When no better points are generated in the trust region, the radius of the trust region $\rho$ decreases and then continues to generate new points as above. The above process is repeated until $\rho$ reaches minimum tolerance or maximum iteration is reached.

\section{Results and discussion}

In this paper, we use the low energy storage ring of Wuhan Photo Source (WHPS)\cite{bib:52} as an example. WHPS is a fourth-generation light source that is currently being prepared for construction. It comprises a low-energy storage ring (1.5 GeV), a medium-energy storage ring (4.0 GeV), and a linac that serves as a full-energy injector. The low energy storage ring has 8 cells, a circumference of 180 m, and an emittance of 238.4 pm rad.  

In this paper, short-term tracking only tracks one turn. As the number of qubits increases, simulations of quantum computing on a classical computer will have a dramatic increase in computational resources as well as computational time. Therefore, for a particle, we select only the part of its trajectory as a feature instead of the complete trajectory. The dynamic aperture refers to the limitation of particle motion in the transverse direction. Hence, we only consider the variables associated with the transverse direction as features, namely $[x,x',y,y']$. The dataset that uses $[x,x',y,y']$ of the end point of the last cell of the low energy storage ring as the features is referred to as the \textit{simple dataset}. On the other hand, the dataset that uses $[x,x',y,y']$ of the end points of the first, fifth, and last cells as the features is referred to as the \textit{complex dataset}. The number of features contained in each data sample of these two datasets is 4, 12, respectively. Jiao et al. assign a higher sampling density to particles with larger amplitudes when generating initial particles. Based on their method, we initially made the distance between these semicircles decay exponentially from inside to outside. However, we found that this resulted in far more particles being lost than survived for long-term tracking, which led to an imbalance in the ratio of labels in the dataset. Therefore, as shown in Fig.~\ref{fig:nine-column-figure} we divide the particles into three intervals from inside to outside, with the same density of particles in each interval and increasing density in order, making the number of surviving particles close to that of lost particles, where the number of surviving particles is 2546 and the number of lost particles is 2454.

When the computational resources are set to 20 Xeon 2.4 GHz CPUs in the Supercomputing Center of Wuhan University cluster and the training dataset consists of 500 samples, the training time for VQC on the ideal simulator is approximately 16 minutes for the simple dataset, while ANN completes training within seconds. Moreover, when using the complex dataset, the training time for VQC is around 175 minutes, whereas ANN still takes only seconds. Additionally, as the training dataset size increases, the required time for VQC also significantly increases. However, once the training is complete, the prediction time of VQC can be controlled within a few seconds.

\begin{figure}[!htb]
\includegraphics
  [width=0.6\hsize]
  {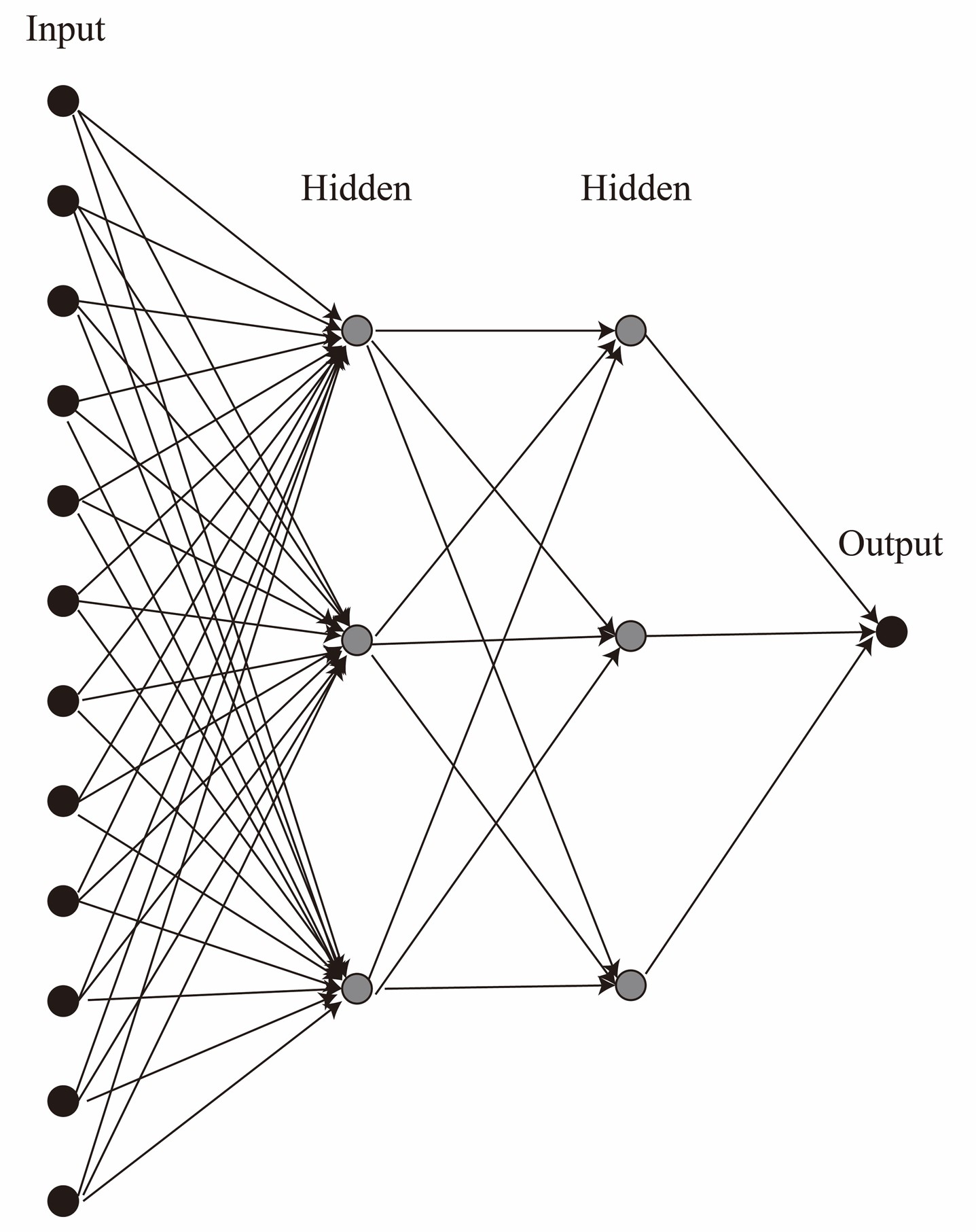}
\caption{The ANN model diagram, with two hidden layers, each containing three nodes, resulting in a total of 108 trainable variables.}
\label{fig:six-column-figure}
\end{figure}

\subsection{Ideal simulation}
The simulation time of a quantum circuit exhibits a steep increase as the number of quantum gates grows, while for a QML model, its performance is determined by the number of quantum gates (i.e., the number of trainable variables) it contains. The selection of an appropriate number of strong entangling layers is crucial for balancing the speed and performance of VQC. In this study, we investigate this trade-off by employing the \textit{simple dataset} and angle encoding. Specifically, the VQC model is trained using different depths of strong entangling layers, ranging from 1 to 7, and their performance is subsequently compared. To accelerate the experimental process, the training dataset is restricted to 1000 samples. Due to the relatively small size of the training set in this study, only one batch is used for training, and the maximum number of iterations is set to 1000. The model is simulated using the AerSimulator provided by Qiskit without any noise for 1024 measurements of the final state of this circuit. The same quantum model simulation setup is employed for all experiments in the ideal simulation section. Finally, the average of the accuracy obtained by ten independent training sessions with different test sets is used as the metric to evaluate the model performance. This metric will also be utilized for subsequent experiments.

\begin{figure}[!htb]
\includegraphics
  [width=0.9\hsize]
  {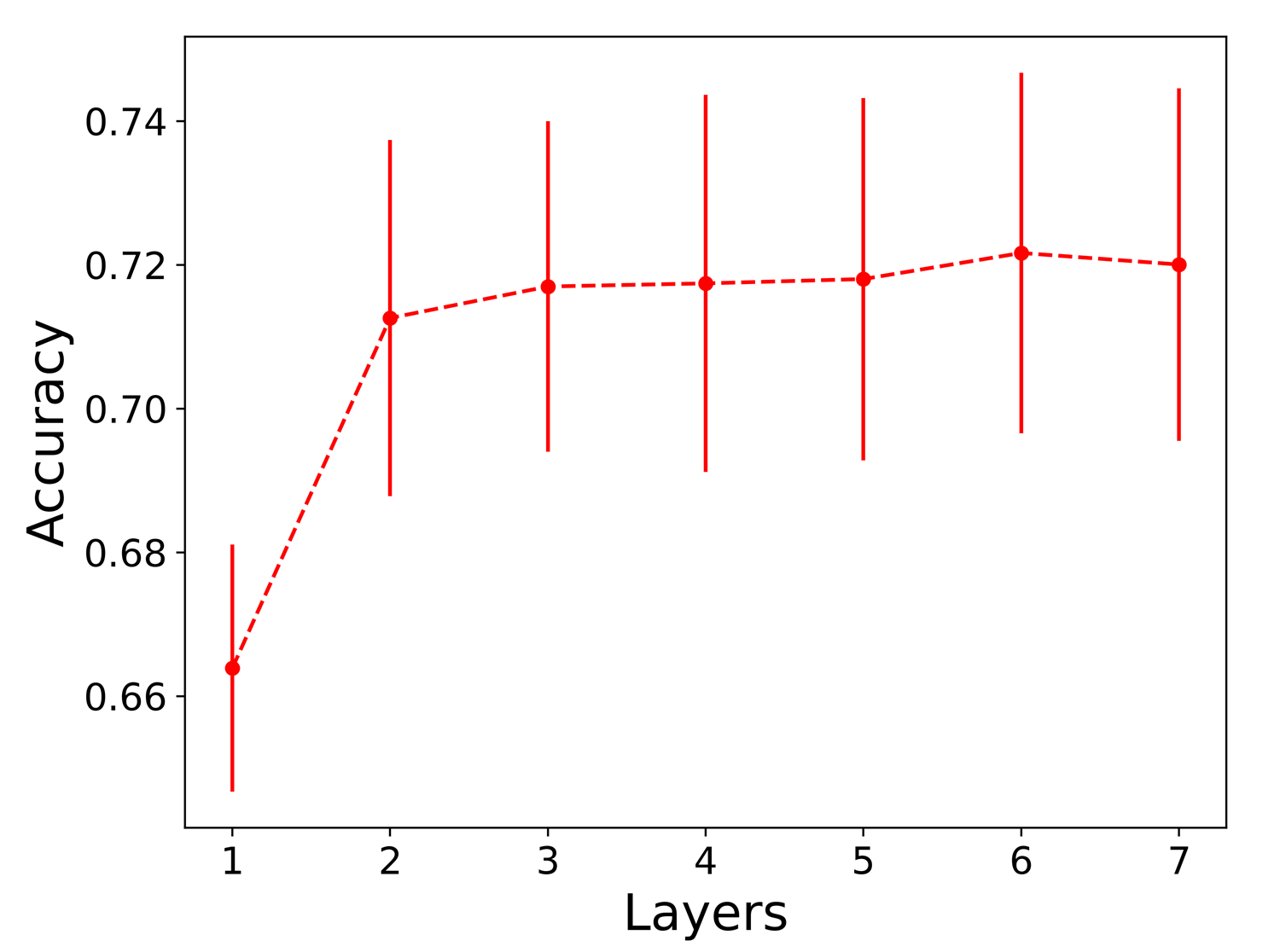}
\caption{Accuracy of VQC with angle encoding on the simple dataset versus the number of strongly entangled layers.}
\label{fig:seven-column-figure}
\end{figure}

Fig.~\ref{fig:seven-column-figure} shows the accuracy trend of VQC using angle encoding as the number of strong entangling layers increases. It is expected that more layers enhance the complexity and improve accuracy. However, after layer 2, the accuracy improvement becomes significantly slower. This suggests that adding more layers does not significantly enhance VQC performance, despite longer simulation time. Therefore, using a small number of strong entangling layers achieves good performance while reducing computational time. Consequently, all subsequent experiments will use a VQC model with only three strong entangling layers.

After analyzing the influence of the number of strong entangling layers on the VQC model, a comparative study will be conducted to examine the performance between VQC models employing different encoding methods and classical ANN. Due to the limited number of features in the simple dataset, the information that can be extracted by VQC may be insufficient, resulting in low accuracy levels as depicted in Fig.~\ref{fig:seven-column-figure} To achieve higher accuracy that is more aligned with practical requirements and enhance the meaningful comparison between different methods, the \textit{complex dataset} is utilized in the comparative experiments and the maximum number of iterations is set to 3000. The VQC model used for amplitude encoding involves replacing the encoding scheme depicted in Fig.~\ref{fig:three-column-figure} with the amplitude encoding. In this experimental setup, angle encoding requires 12 qubits, while amplitude encoding only requires 4 qubits. The ML algorithm used to compare with it is the ANN model on scikit-learn\cite{bib:53}. To ensure a fair comparison between the two models, we refer to previous studies\cite{bib:20, bib:21, bib:28} and use the similar number of training parameters for both models. The ANN model as shown in Fig.$~\ref{fig:six-column-figure}$. Due to the identical VQC model architecture employed for both angle encoding and amplitude encoding, the number of qubits used for amplitude encoding is only one-third of that used for angle encoding. As a result, the number of training parameters utilized in amplitude encoding is also one-third of that used in angle encoding. Taking into account the aforementioned factors, the number of training parameters is set to 36, 108, and 108 for angle encoding, amplitude encoding, and ANN, respectively. 

\begin{figure}[!htb]
\includegraphics
  [width=0.9\hsize]
  {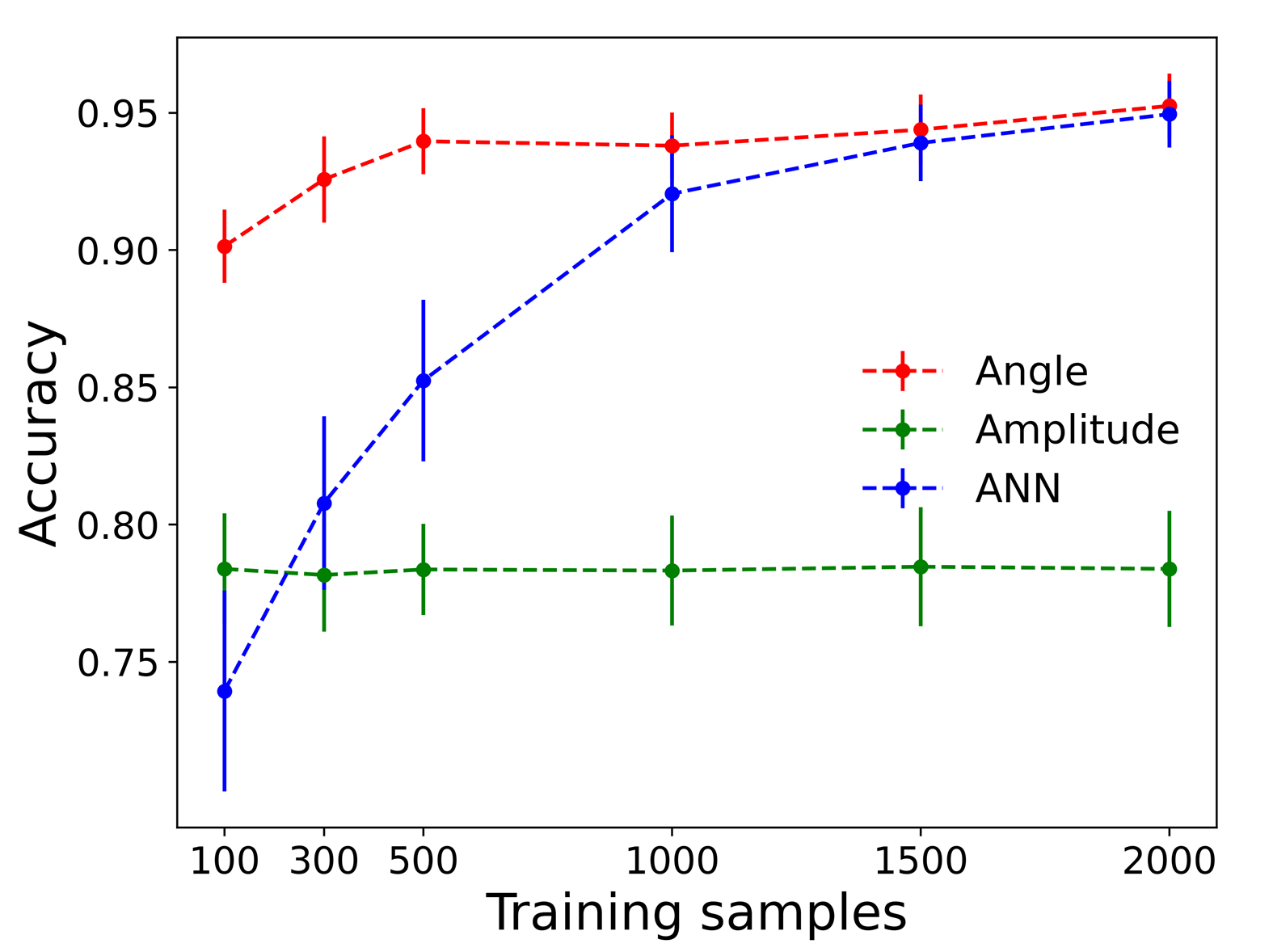}
\caption{Accuracy of angle encoding (red), amplitude encoding (green), and ANN (blue) on the complex dataset versus the number of training samples.}
\label{fig:eight-column-figure}
\end{figure}

Fig.~\ref{fig:eight-column-figure} shows the variation in the accuracy of the three models as the training samples increase. The VQC models corresponding to the two different encoding schemes exhibit a faster convergence to their optimal accuracies with an increasing number of training samples. For the amplitude encoding, the accuracy tends to plateau after 300 training samples, showing limited improvement beyond this point. This phenomenon has also been observed in other studies\cite{bib:21, bib:57}. The current researches suggest that it is not a random occurrence but likely represents an advantage of QML. The demonstration of this advantage warrants further investigation and research. Most state-of-the-art ML algorithms typically require large datasets to achieve high performance, while VQC appears to be less sensitive to dataset size. Since collecting large datasets in real-world applications can often be time-consuming and expensive, this is particularly meaningful and deserves further study.

It is worth noting that when the number of training samples is below 500, angle encoding performs better than ANN. This observation aligns with a proved property of the QML models \cite{bib:57}. The ultimate goal of ML is to accurately predict unknown data, which is also referred to as generalization. The generalization performance of a QML model with training parameters denoted as $\alpha$ can be quantified using the measure of generalization error, represented as \textit{${gen(\alpha)}$}. A smaller value of \textit{${gen(\alpha)}$} indicates better generalization capability of the model. Suppose there are N training samples and a QML model with T independently parameterized gates, and $\alpha'$ represents the final parameters of the trained model. The generalization error of the trained model satisfies

\begin{equation}\label{eq:10}
\textit{\text{gen}}(\alpha') \in O\left(\sqrt{\frac{T \log T}{N}}\right)
\end{equation}

from this, it can be inferred that ${gen(\alpha')}$ increases as T/N increases. In classical statistical learning theory, model complexity is often measured by the Vapnik-Chervonenkis(VC) dimension$\cite{bib:54}$, denoted as $d$. The generalization error typically follows $O\left( \sqrt{\frac{d}{N}} \right)$$\cite{bib:54, bib:55}$, for complex models such as neural networks or nonlinear decision trees, the model complexity $d$ is generally large. Thus, when $N$ is relatively small (for example, less than 2000), VQC can demonstrate better performance. This aligns with the findings in Fig.7.

\begin{figure}[!htb]
\includegraphics
  [width=1\hsize]
  {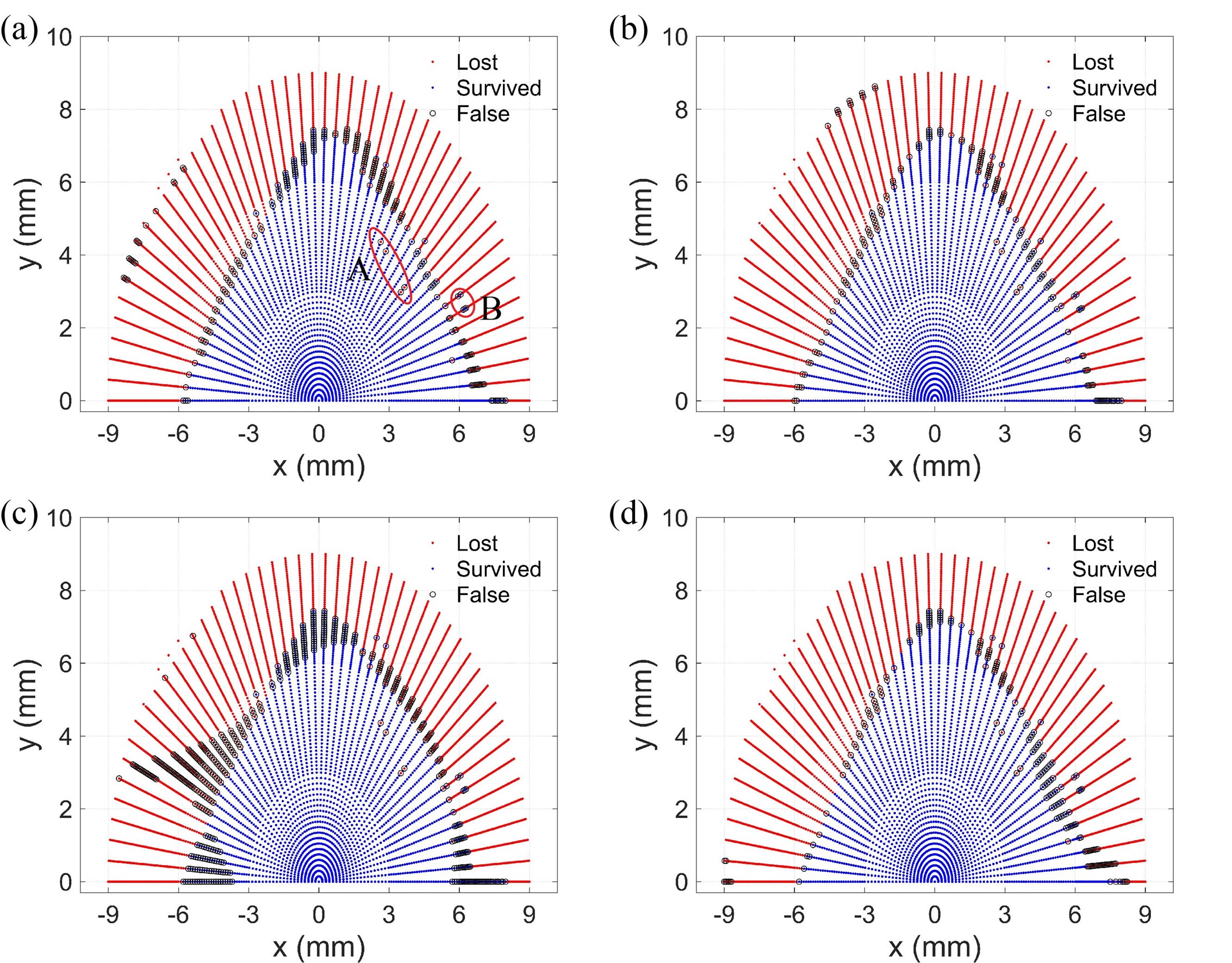}
\caption{The results of VQC and ANN trained with different training samples to predict dynamic aperture. Red and blue states represent the particle survival states obtained through long-term tracking. Black circles indicate cases where the model predicts incorrectly. \textbf{a} and \textbf{b} show the results of VQC with training samples of 500 and 2000, respectively, while \textbf{c} and \textbf{d} show the corresponding results of ANN.}
\label{fig:nine-column-figure}
\end{figure}

The accuracy of amplitude encoding consistently remains lower than that of angle encoding, which is consistent with the findings reported in a previous study\cite{bib:21}. Therefore, the subsequent comparisons and experiments in VQC refer to models that utilize angle encoding. It is evident that the accuracy of VQC is consistently higher than that of ANN for different training sample sizes as shown in Fig.~\ref{fig:eight-column-figure}. For instance, when the training sample size is 2000, the accuracy of VQC reaches 95.3\%, while the accuracy of ANN is 94.9\%. The prediction of dynamic aperture by VQC and ANN trained with 500 training samples and 2000 training samples is shown in Fig.~\ref{fig:nine-column-figure}. A portion of the particles is missing in the upper left corner of the semi-circular region where the initial particles are located. This is because this section of the particles was lost during the initial short-term tracking, and therefore, a complete trajectory was not obtained. Therefore, this portion of the particles is not used in the training process. As shown in Fig.~\ref{fig:nine-column-figure}(a), four particles in region A surrounded by red contours show loss by long-term tracking, while the particles near them are all alive. Conversely, in region B, several particles demonstrate the opposite phenomenon. This is because tracking the dynamic aperture is a complex nonlinear problem, and occasionally individual particles may exhibit this behavior. Given this complexity and uncertainty, this prediction method is more suitable for ring-type electron accelerators with radiation damping. Based on the figure, it is evident that neither VQC nor ANN can accurately predict the outcome in this scenario. 

\subsection{Noisy models}
\begin{figure}[!htb]
\includegraphics
  [width=1\hsize]
  {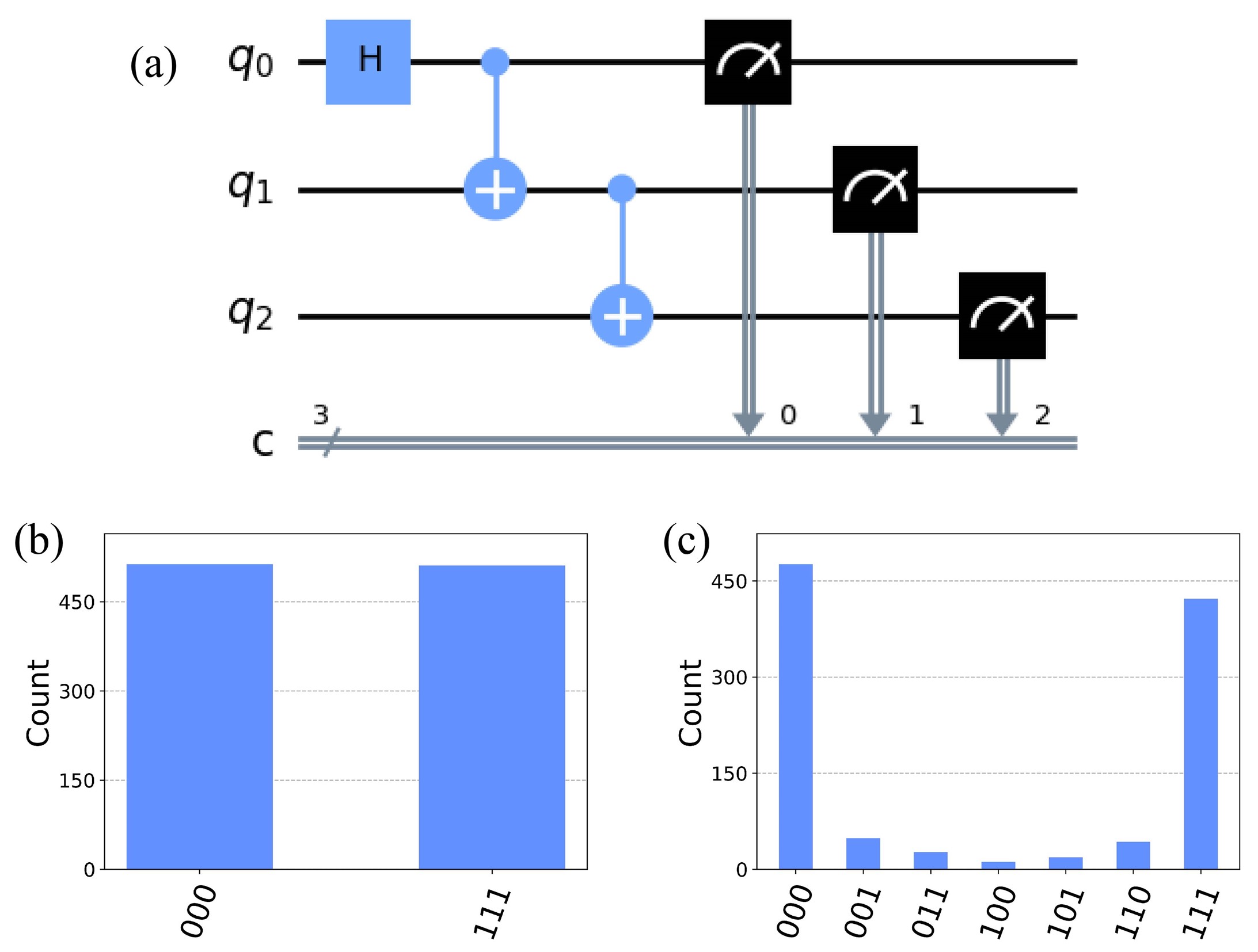}
\caption{Comparison of the results of a 3-qubits circuit on an ideal simulator and a real quantum computer.\\
\textbf{(a)} The circuit diagram of a 3-qubit quantum system. $q_0$, $q_1$, and $q_2$ represent the quantum bits (qubits), and $c$ represents the classical bit. $H$ is the Hadamard gate, which is used to create a superposition state on the qubits.\\
\textbf{(b)} and \textbf{(c)} The measurement results of the quantum circuit on an ideal simulator versus a real quantum computer. The horizontal axis shows the possible outcomes of the 3-qubit quantum state, labeled as $000$ and $111$, The vertical axis represents the count, or the frequency of each outcome occurring after multiple runs of the quantum circuit, with a total of 1,024 shots.}
\label{fig:ten-column-figure}
\end{figure}

The majority of current real quantum computers are NISQs and the results obtained from these machines are slightly different from those obtained from noiseless simulators. To visualize the difference between an ideal simulator and a real quantum computer, a 3-qubit circuit consisting of one H gate and two CNOT gates, as shown in Fig.~\ref{fig:ten-column-figure}(a), is created and run using the ideal simulator on the AerSimulator provided by Qiskit and an IBM quantum computer, ibmq\_belem, with a measurement of 1024. In theory, the output of this circuit has only two states, $\left| 000 \right\rangle$ and $\left| 111 \right\rangle$, each with a probability of 0.5. The output of the ideal simulator, as shown in Fig.~\ref{fig:ten-column-figure}(b), is consistent with the theory, with only the expected two states and a probability close to 0.5. However, the output of the real quantum computer, as shown in Fig.~\ref{fig:ten-column-figure}(c), has five additional states, which deviate from the theoretical result. It is worth noting that the state |010⟩ is not observed in Fig.~\ref{fig:ten-column-figure}(c) because, apart from the states $\left| 000 \right\rangle$ and $\left| 111 \right\rangle$, all other states are generated due to noise, resulting in very low probabilities of occurrence. As a result, $\left| 010 \right\rangle$ did not appear in the limited number of measurements conducted. Therefore, it is essential to study the performance of VQC in the presence of noise.

In NISQ hardware, several errors can cause the output to deviate from the ideal simulator, including the gate error probability of each basis gate on each qubit, the gate length of each basis gate on each qubit, the T1 and T2 relaxation time constants of each qubit, and the readout error probability of each qubit. Qiskit-Aer, a provider of Qiskit, can simulate the errors present in the IBMQ real quantum computers, which are used in this paper for noise studies. Three IBM quantum computers with different structures are used in this study: ibmq\_lima, ibmq\_belem, and ibm\_lagos, which have 5, 5, and 7 qubits, respectively, and corresponding Quantum Volumes\cite{bib:58} of 8, 16, and 16.

In this part, the simple dataset is used due to the high computational resources and time required for simulating the noise model. As shown in Fig.~\ref{fig:eight-column-figure}, VQC can achieve good performance even with smaller training samples, and hence, the number of training samples used is set to 1000. The VQC model employs angle encoding and three strongly entangled layers with COBYLA as the optimizer, with a maximum number of iterations set to 1000. According to the results presented in Table~\ref{tab1}, it can be observed that the use of the noise model in VQC only causes a very slight decrease in prediction accuracy compared to the ideal model. The difference between them is not significant. Moreover, it is evident that the prediction accuracy of VQC corresponding to the \textit{simple dataset} is much lower than that corresponding to the \textit{complex dataset}. This can be attributed to two reasons: firstly, the number of features in the complex dataset is much larger than that in the \textit{simple dataset}. Secondly, the number of training parameters for the VQC model when using the complex dataset is also much larger.

\begin{table}[!htb]
\caption{Accuracy for noisy models with the simple dataset.}
\label{tab1}
\setlength{\tabcolsep}{4pt} 
\scalebox{0.8}{
\begin{tabular}{lllll}
\toprule
Models & no noise & ibmq\_lima & ibmq\_belem & ibm\_lagos \\
\midrule
Accuracy  & 0.717 ± 0.223 & 0.688 ± 0.022 & 0.686 ± 0.028 & 0.692 ± 0.018 \\
\bottomrule
\end{tabular}
}
\end{table}

As shown in Equation~\eqref{eq:8}, VQC uses the probability of being greater than 0.5 when the output state is $\left| 0 \right\rangle$ to determine the label, which does not require a very precise value. As seen in Fig.~\ref{fig:ten-column-figure}(c), the change in the output state in the presence of noise is not significant. This makes it most likely that the relationship between the probability of the VQC output state being $\left| 0 \right\rangle$and the magnitude of 0.5 remains unchanged, resulting in only a slight drop in VQC performance in the presence of noise, but it still performs well.

\section{Conclusion}

QML algorithms combine quantum computing with classical ML algorithms and have great potential for improving classical ML algorithms due to the advantages of quantum computing. In this paper, the application of QML algorithms to evaluate dynamic aperture in accelerator physics has been presented. The concepts of QML and VQC have been introduced. Quantum circuit simulation requires significant computational resources, while COBYLA requires significantly fewer simulations per iteration compared to gradient-based optimization algorithms. Therefore, we have used the COBYLA algorithm to train the VQC model to reduce the demand for computational resources. 

The relationship between the performance of VQC and the number of the strong entangling layers has been studied. It has been observed that beyond a certain number of layers, increasing the number of layers no longer leads to an improvement in the performance of VQC. Experimental results indicate that the performance of amplitude encoding is inferior to angle encoding. The performance of VQC and ANN with the same number of training parameters has been compared. When the training sample size is relatively small compared to the training set, VQC performs significantly better than ANN and can almost accurately predict the dynamic aperture. The ability of VQC to achieve good results with a relatively small number of training samples is very meaningful for problems where obtaining training samples is very expensive. As a more concrete example, researchers typically employ Bayesian optimization algorithms for online tuning of particle accelerators\cite{bib:59, bib:60, bib:61}. In such scenarios, our QML approach can effectively train neural network models with comparatively small datasets and integrate the NN into the Bayesian optimization iteration process. As the number of training samples increases, the accuracy of VQC and ANN becomes very close, with VQC achieving an accuracy of 95.3\% and ANN achieving an accuracy of 94.9\%. The impact of noise on quantum computing results has been demonstrated by comparing the performance of a circuit on an ideal model and a real quantum computer. The accuracy of VQC using noisy models only decreased slightly compared to an ideal model. This indicates that the VQC is robust against noise and can be applied to recent NISQs. This study provides a novel perspective for researchers in the field of accelerator physics and offers valuable insights for those interested in exploring the application of QML in various domains.

\section{Declaration of competing interest}

The authors declare that they have no known competing financial interests or personal relationships that could have appeared to influence the work reported in this paper.

\section{Data availability}

Data will be made available on request.

\section{Acknowledgments}

We thank the Supercomputing Center of Wuhan University. The numerical calculations in this paper have been done on the supercomputing system in the Supercomputing Center of Wuhan University.

\end{document}